\documentclass[12pt,preprint]{aastex}

\def\kms{km s$^{-1}$}
\def\vrot{$v_{e}$sin i}

\usepackage{lscape}
\begin{document}

\title{Spinning like a Blue Straggler: the population of fast rotating Blue Straggler stars
in $\omega$ Centauri
\footnote{Based on observations collected at the ESO-VLT under the programs 
077.D-0696(A), 081.D-0356(A) and 089.D-0298(A).}
}

\author{A. Mucciarelli$^{1}$, L. Lovisi$^{1}$, 
F. R. Ferraro$^{1}$, E. Dalessandro$^{1}$, B. Lanzoni$^{1}$, 
L. Monaco$^{2}$}

\affil{$^{1}$Dipartimento di Fisica \& Astronomia, Universit\`a 
degli Studi di Bologna, Viale Berti Pichat, 6/2 - 40127
Bologna, ITALY}

\affil{$^{2}$European Southern Observatory, Casilla 19001, Santiago, Chile}

\begin{abstract}
By using high-resolution spectra acquired with FLAMES-GIRAFFE at the
ESO/VLT, we measured radial and rotational velocities for 110 Blue
Straggler stars (BSSs) in $\omega$ Centauri, the globular cluster-like
stellar system harboring the largest known BSS population.  According
to their radial velocities, 109 BSSs are members of the system.  The
rotational velocity distribution is very broad, with the bulk of BSSs
spinning at less than $\sim$40 \kms\ (in agreement with the majority
of such stars observed in other globular clusters) and a long tail
reaching $\sim$200 \kms.  About 40\% of the sample has \vrot\ $>$40
\kms\ and about 20\% has \vrot\ $>$70 \kms.  Such a large
fraction is very similar to the percentage of of fast rotating BSSs
observed in M4. Thus, $\omega$ Centauri is the second stellar cluster,
beyond M4, with a surprisingly high population of fast spinning BSSs. 
We found a hint of a radial behaviour of the fraction of fast rotating 
BSSs, with a mild peak within one core
  radius, and a possibile rise in the external regions (beyond four
  core radii).  This may suggest that recent formation episodes of
  mass transfer BSSs occurred preferentially in the outskirts of
  $\omega$ Centauri, or that braking mechanisms able to slow down
  these stars are least efficient in lowest density environments.
\end{abstract}


\section{Introduction}
Blue Straggler Stars (BSSs) are nowadays firmly established as an
"anomalous" class of stars populating any stellar
environment, ranging from open star clusters \citep{js55,mg09}, to
globular clusters \citep{san53,fer92,fer03_six,pio04,lei07}, the
Galactic field \citep{ps00}, and dwarf spheroidal galaxies
\citep{momany07,mape09}.

In the optical color-magnitude diagrams (CMDs) BSSs appear as objects
brighter and bluer (hotter) than the normal turnoff (TO) stars, lying
along an extension of the main-sequence (MS).  Their position in the
CMD \citep{f06_tuc,lan07_19} and the direct spectroscopic and
pulsation measurements \citep{shara97,demarco05,gilligand98,fiore14},
have shown that BSSs are more massive than normal TO stars.  Two main
scenarios have been proposed for their formation: direct
  stellar collisions \citep[COL-BSSs;][]{hills76}
  and mass transfer activity between binary
  companions \citep[MT-BSSs;][]{mccrea64}, either due to stellar
  evolution or triggered by stellar interactions, possibly up to the
  complete coalescence of the two objects.  The two mechanisms can
possibly occur simultaneously in dense stellar systems, like globular
clusters \citep[GCs; ][]{fer93,fer97,fer09_double,dale13}.  Hence,
these objects not only are the tangible proof of the (mild or violent)
interactions occurring between stars in GCs, but they also offer the
opportunity of investigating the internal dynamics of stellar systems
\citep{fer01,mape04,mape06,lan07_m5,lan07_6388,dale08_6388,sabbi04,bec06}. In
this respect one of the most recent exciting results has been obtained
by \citet{fer12}, who demonstrated that the BSS radial distribution
can be used as a clock to measure the parent cluster dynamical age.

In spite of their importance, many issues concerning the BSS formation
and properties are still open. In particular, one of the most
challenging problems is the identification of observable features able
to discriminate between the two formation channels.  Negligible mixing
between the inner cores and the outer envelopes is predicted for
COL-BSSs \citep{lombardi} that should show normal C and
O abundances.  Conversely, MT-BSSs are expected to show C and
O depletion on their surface \citep{sarna}, because the material
should come from the inner regions of the donor star where the
CNO-cycle occurred.  Indeed this chemical signature has been observed
in a sub-sample of BSSs in 47 Tuc \citep{f06_tuc} and in M30 (Lovisi
et al 2013a). Another interesting measurable BSS characteristic is the
projected rotational velocity, \vrot.  Unfortunately, in this
case the theoretical interpretative scenario is significantly more
complex since both MT- and COL-BSSs are expected to rotate fast
\citep{sarna,benz87}, but braking mechanisms are suggested to occur
and slow down the stars, with timescales and efficiencies which are
still unknown \citep{leonard95, sills05}.

From the observational point of view, most of the BSSs appear to be
slow rotators (with rotation velocities lower than 40 \kms) in all the
GCs studied so far by our group, namely 47 Tuc \citep{f06_tuc}, NGC 6397
\citep{lovisi10}, M30 \citep{lovisi13a} and NGC 6752
\citep{lovisi13b}.
Even though the range of rotational velocities is larger than that
spanned by normal MS stars in GCs \citep[see e.g.][]{lucatello03}, it
is still compatible with that of the fastest rotating stars in
"normal" evolutionary sequences in GCs, namely horizontal branch stars
redder than the Grundahl jump \citep{grund} with 8000 K$< T_{\rm
  eff}<$12000 K \citep[see][]{peterson95, behr00a,
  behr00b}. Nevertheless, some outliers have been found in 47 Tuc,
M30 and NGC 6397 (but not in NGC~6752): one BSS per cluster shows very
high rotation, up to more than 90 \kms. However, the most
surprising result has been found for M4: out of the 20 BSSs
investigated in this GC, \citet{lovisi10} identified 8 fast-rotating
(FR) BSSs (with \vrot\ $>$40 \kms), corresponding to 40\%
of the studied sample. Indeed, this is the largest percentage of FR
BSSs ever found in any GC.

Here we present the results obtained for a sample of BSSs in the
stellar system $\omega$ Centauri.  All the evidence collected so far
\citep{lee99,pancino00,fer04_ome,nd95,sollima05,ori03,jp10} suggest
that $\omega$ Centauri is not a {\sl genuine} GC\footnote{Note that
  only another GC-like stellar system with a similarly large
  metallicity spread has been found in the Galaxy: Terzan 5 in the
  Galactic Bulge \citep{fer09_ter,ori11,ori13}.}, but possibly the
remnant core of a tidally disrupted dwarf galaxy \citep[see
  e.g.][]{dinescu99,maje00,bekki06}. This stellar system hosts the
richest BSS population observed so far in a GC, with more than 300
candidates \citep{f06_ome}.  Its normalized BSS radial distribution is
flat, suggesting that the system is not dynamically relaxed yet
\citep{fer12};  note that a similar distribution has been found in
  only two other stellar systems: NGC 2419 \citep{dale08_2419} and Pal
  14 \citep{bec11}.  This means that mass segregation had not enough
time yet to be effective and stellar interactions are very
unfrequent.  For these reasons the BSS population of $\omega$
  Centauri is thought to have mainly formed through MT activity in
  primordial binary systems \citep{f06_ome}. This makes $\omega$
Centauri the ideal stellar system for investigating the properties of
MT-BSSs.

Recently, \citet[][hereafter SP14]{simunp} measured rotational
velocities for 49 BSSs in $\omega$ Centauri by using medium-resolution
spectra (R$\sim$10000) taken with the Inamori Magellan Areal Camera
and Spectrograph (IMACS) multimode facility at the Baade Magellan
telescope. They found a quite large distribution of \vrot, peaked
around $\sim$ 20-30 \kms\ and reaching about 170 \kms.  One of the
most intriguing results of their work is a hint that the FR BSSs are
more centrally concentrated with respect to slower BSSs.  Moreover, a
trend between rotational velocity and color (hence, temperature) has
been claimed. Based on these results the authors suggested that the FR
BSSs formed preferentially in the central regions and are also more
massive than slow rotators.
 
As part of a systematic spectroscopic campaign aimed at studying the
chemical and kinematicl properties of BSSs in GCs (see previous results
in Ferraro et al. 2006a, Lovisi et al. 2010, 2013a,b), in this paper
we present and discuss the rotation velocity distribution of 110 candidate BSSs
in $\omega$ Centauri.
 
The paper is organized as follows: Section 2 describes the observations, 
followed in Section 3 by the derivation of the atmospheric parameters. 
The measurements of radial and rotational velocities are discussed in Section 4 and 5, 
respectively. Section 6 discusses the comparison between our results and those of SP14. 
Finally, Section 7 presents the discussion of the results and our conclusions.

\section{Observations}
The spectroscopic dataset analysed in this paper includes two samples
of high-resolution spectra acquired with the multi-object spectrograph
FLAMES-GIRAFFE \citep{pasquini} mounted at the Very Large Telescope of
the European Southern Observatory (ESO):
 
\begin{enumerate} 
\item {\sl Dataset 1} --- 73 BSSs observed with the gratings HR5A
  (R$\sim$18500, $\Delta\lambda = 4340-4587 \mathring{A}$), HR14A
  (R$\sim$18000, $\Delta\lambda = 6308-6701 \mathring{A}$) and HR15N
  (R$\sim$17000, $\Delta\lambda = 6470-6790 \mathring{A}$), during
  the ESO periods 81 and 89 (PI: Ferraro).  For each target 3
  exposures of $\sim$3600 s with both HR14A and HR15N gratings, and 5
  exposures of $\sim$3000 sec with the HR5A have been secured.

The targets have been selected from a photometric catalog obtained by
combining high resolution images (through the $B$ and $R$ filters)
acquired with the ACS camera on board of HST and wide-field images
obtained with the ESO-WFI imager (through the $B$ and $I$ bands; see
\citealp{f06_ome} for the description of the photometric dataset).  We
conservatively excluded targets having close stellar sources of
comparable or higher luminosity within $3\arcsec$, in order to avoid
contamination of the spectra from close companion stars. 

\item {\sl Dataset 2} --- 54 BSSs observed with the gratings HR2
  (R$\sim$19600, $\Delta\lambda = 3854-4049 \mathring{A}$) and HR4
  (R$\sim$20350, $\Delta\lambda= 4340-4587 \mathring{A}$) and
  retrieved from the ESO Archive (ESO period 77, PI: Freyhammer).  8
  and 6 exposures of $\sim$1250 s each have been secured with the
  gratings HR2 and HR4, respectively.  The coordinates of the targets
  have been cross-correlated with our photometric catalog, in order to
  put all the targets in the same astrometric and photometric systems.
  We checked that all the targets are not contaminated by bright
  companions.  A total of 17 BSSs are in common between the two
  datasets.
\end{enumerate}

\begin{figure}
\plotone{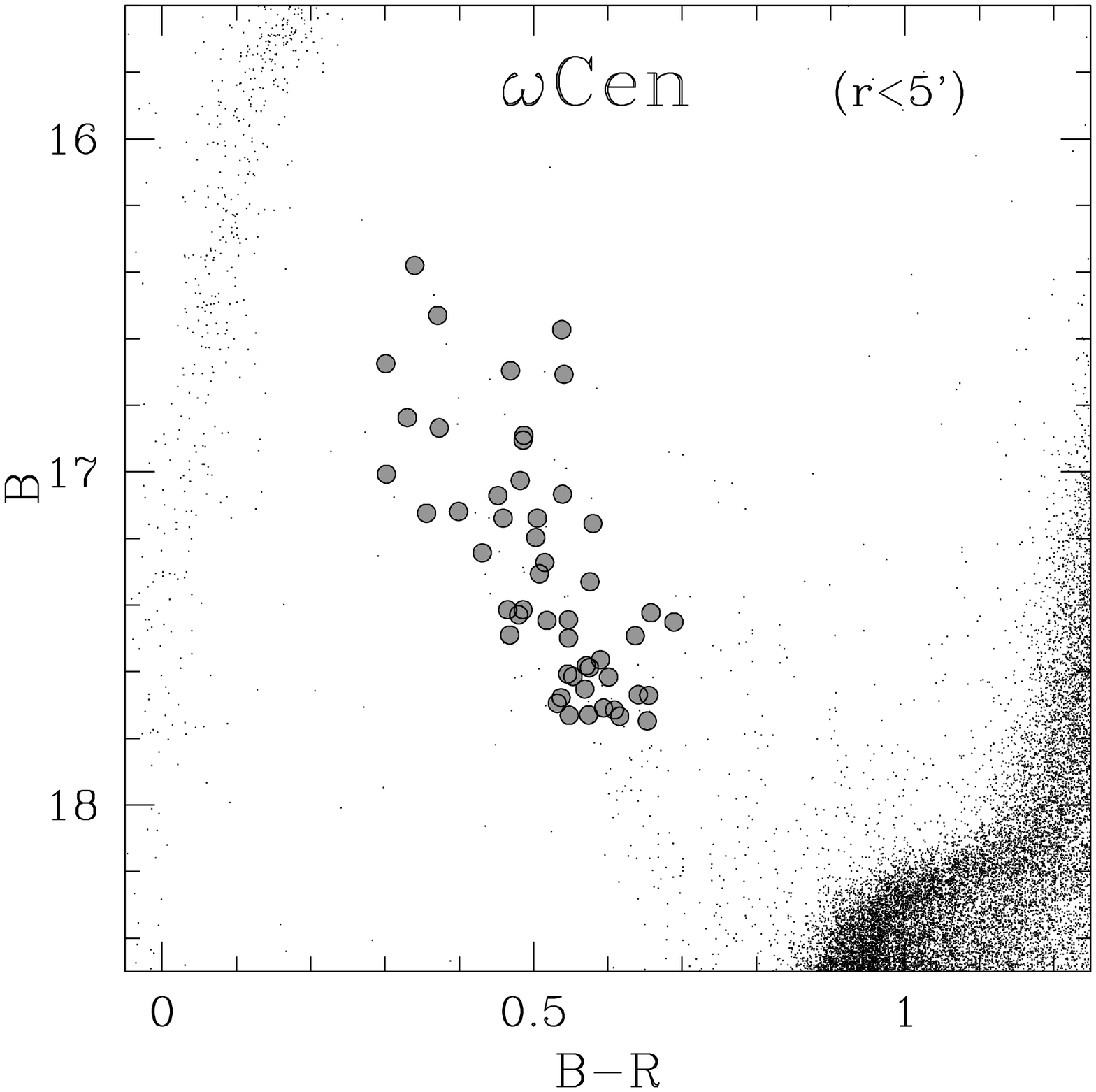}
\caption{ACS/HST CMD of the central region ($r<5\arcmin$) of $\omega$
  Centauri zoomed in the BSS region. The spectroscopic targets
  discussed in the paper are marked with large grey circles.}
\label{cmd1}
\end{figure}

\begin{figure}
\plotone{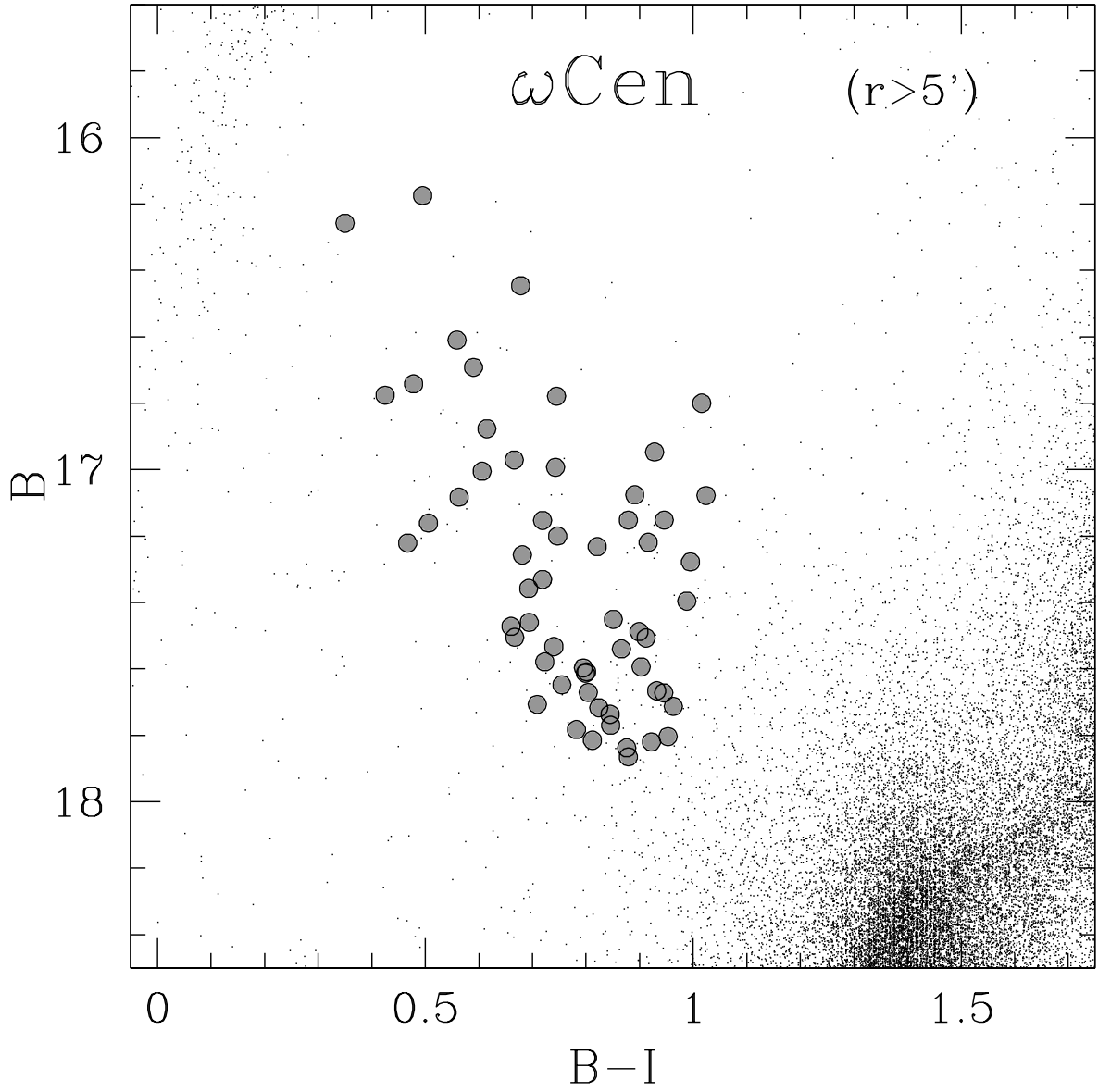}
\caption{As in Fig. \ref{cmd1}, but for the external region
  ($r>5\arcmin$) of $\omega$ Centauri. }
\label{cmd2}
\end{figure}

The position of the observed targets in the CMDs is shown in Figures
\ref{cmd1} and \ref{cmd2}.  Both the datasets have been reduced with
the version 2.9.2 of the GIRAFFE ESO
pipeline\footnote{http://www.eso.org/sci//software/pipelines/}
including bias-subtraction, flat-fielding, wavelength calibration and
spectral extraction.  For each grating, the sky contribution has been
removed from each individual exposure by using a master-sky spectrum
obtained as the median of all the sky spectra.  Finally, for any given
star the sky-subtracted spectra (corrected for radial velocity, RV;
see Section \ref{rv}) have been coadded together.

\section{Atmospheric parameters and synthetic spectra}

Atmospheric parameters (effective temperature and surface gravity) 
for our targets have been derived
photometrically, by orthogonally projecting the stellar position in
the CMDs on a set of theoretical isochrones with different ages (from
100 Myr to 5 Gyr), able to cover the entire extension of the BSS
sequence in the CMD. In particular, we used isochrones from the Padova
database
\citep{Bressan12}\footnote{http://stev.oapd.inaf.it/cgi-bin/cmd},
with metallicity Z=0.0006 and $\alpha$-enhanced chemical mixture
(corresponding to [M/H]$\sim -1.5$ dex), and we assumed a distance
modulus (m-M)$_0=13.7$ mag and a color excess E$(B-V)=0.11$ mag 
\citep{f06_ome}, suitable to best-fit the old, metal-poor ([Fe/H]$\sim
-1.5$ dex; see e.g. \citealp{jp10}) component of $\omega$ Centauri with
a 13 Gyr-isochrone.  A microturbulent velocity of 0 \kms\  is
adopted for all the targets, because of the shallow convective (or
fully radiative) envelopes expected for these stars (note that
different assumptions for the microturbulent velocity have no impact
in the determination of RVs and \vrot).

In order to measure RVs and \vrot\  we made extensive use of
synthetic spectra computed with the code {\tt SYNTHE}
\citep{sbordone04,kurucz05}, adopting the last version of the
Kurucz/Castelli linelist for atomic and molecular transitions. The
line-blanketed model atmospheres have been computed with the code {\tt
  ATLAS9} \citep{castelli04}, under the assumptions of Local
Thermodynamic Equilibrium for all the species, one-dimensional,
plane-parallel geometry and without the inclusion of the approximate
overshooting in the flux computation. For each star an ATLAS9 model
atmosphere has been generated adopting the appropriate stellar
parameters of the target and assuming a metallicity [M/H]=--1.5 dex.

The synthetic spectra have been convolved with a Gaussian profile to
reproduce the instrumental broadening of the different setups. The
broadening for a given setup has been estimated by measuring the FWHM
of bright unsaturated lines in the reference Th-Ar calibration lamp
\citep[following the procedure adopted by][]{behr00a}.  Finally,
rotational velocities have been taken into account by convolving the
synthetic spectra with a rotational profile \citep{kurucz05}.

\section{Radial velocities}
\label{rv}

RVs have been measured through the Fourier cross-correlation technique
as implemented in the IRAF task {\sl fxcor} \citep{tonry}. For each
star we used a synthetic spectrum computed with the appropriate
stellar parameters.
All the templates have been computed by assuming [M/H]$=-1.5$ dex:
note that the assumption of higher metallicities according to the
broad metallicity distribution of $\omega$ Centauri is not a critical
issue and does not affect the measure of RV.

For the targets observed with HR5A, RVs have been derived mainly from
the Mg~II triplet at $4480 \mathring{A}$ (visible in almost all the
spectra observed with this grating) and other transitions when
available.  For the HR4, HR14A and HR15N setups, instead, we used the
Balmer lines.  RVs for the stars observed with the HR2 setup have been
obtained from the photospheric Ca~II K line. We paid special attention
to exclude from the cross-correlation process the Ca~II K interstellar
line visible at $\sim -27$ \kms\  \citep{vanloon09} which is
associated to gas a few kpc far along the line of sight of $\omega$
Centauri and could produce a mismatch in the identification of the
main peak of the cross-correlation function. 

  Multiple RV measurements are available for 74 targets.  We
  carefully verified that the values obtained from different gratings
  and in different epochs are consistent each other. By excluding 17
  BSSs showing hints of RV variations larger than the estimated
  uncertainties (which typically are smaller than 1 \kms), we find
  average differences always smaller than 1 \kms\ 
  between two different gratings, and smaller that 1-2 \kms\ between two
  epochs
  In particular, we found an average difference between Dataset 1 and 
  Dataset 2 of --0.3$\pm$0.6 \kms\  ($\sigma$=1.7 \kms\ ), thus guaranteeing 
  a proper internal accuracy of our measures.  
  By cross-correlating the target list with
  the catalog of variable stars identified by \citet{kal04} and
  \citet{kal05}, we find 27 SX Phe, 6 W UMa, 2 detached eclipsing
  binaries, 1 candidate ellipsoidal variable and 1 variable star with
  a period of $\sim 3.3$ days, but with no classification.  Out of the
  17 BSSs with hints of RV variations, 3 are classified as SX Phe and
  3 as W UMa by \citet{kal04} and \citet{kal05}, while the remaining
  11 do not show photometric variability. These findings demonstrate
  that our dataset is not suitable for studies of RV variability and
  the search for binary systems. We therefore limit to indicate (when
  available) the variable type in Table 1, together with the
  identification name in the catalogs of \citet{kal04} and
  \citet{kal05}. For the targets with multiple measurements we finally
  adopted the average value of RV, assuming as uncertainty the
  dispersion of the mean normalized to the root mean square of the
  number of available measures.

Our sample includes 109 BSS members of the cluster, providing a mean
RV of $233.2 \pm 1.4$ \kms\ ($\sigma =14.7$ \kms). Their coordinates,
atmospheric parameters, and RVs are listed in Table 1.  All but one of
the observed targets have RVs between $\sim 191$ and $\sim 266$ \kms,
in good agreement with the systemic RV of $\omega$ Centauri, peaked at
$\langle$RV$\rangle = 233$ \kms\ \citep[][but see also
  \citealp{mayor97,pancino07,monaco10}]{sollima09}. Only one star
belongs to the field: it has RV$=0.5$ \kms, compatible with that 
  observed \citep[see Figure 7 in ][]{sollima09} and expected
  \citep[see the Galactic model of ][]{besancon} for the thin and
  thick disc stars in the direction of $\omega$ Centauri.

\section{Rotational velocities}
\label{rotv}
Projected rotational velocities have been measured by performing a
$\chi^2$-minimization between the observed spectra and a grid of
synthetic spectra calculated with the atmospheric parameters of each
program star and different values of \vrot.  Note that since abundance
variations affect mainly the line core, while rotation alters the
entire line profile, a correct fit of the entire line profile needs
also the knowledge of the chemical abundances to properly reproduce
the line depth.  Thus, we used an iterative procedure to
simultaneously measure abundance and \vrot, and to provide a reliable
fit of the entire line profile.  Note that for all the stars we
  assumed a global metallicity [M/H]$=-1.5$ dex, leaving the abundance
  of the element under scrutiny to vary during the procedure. We
  checked the impact of this assumption by re-analyzing some stars
  adopting [M/H]$=-1.0$ and $-0.5$ dex (according to the broad
  metallicity distribution of $\omega$ Centauri; see
  \citealp{jp10}). We found variations of \vrot\ of about 0-1 \kms.

\begin{figure}
\plotone{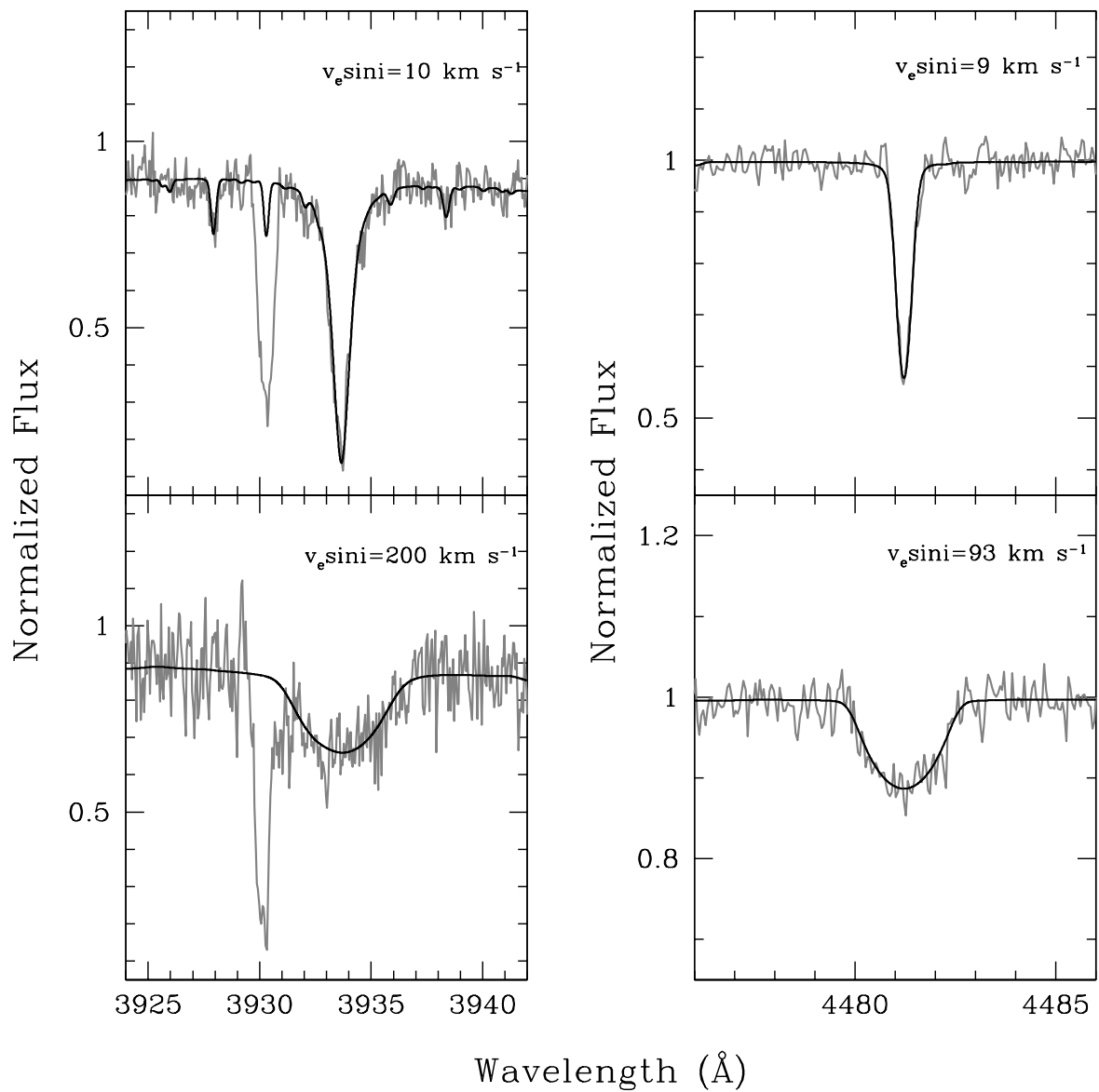}
\caption{Examples of observed spectra (grey lines) with overimposed
  the best-fit synthetic spectra (black lines) computed with the
  rotational velocities labelled.  Left panels show two spectra from
  {\sl Dataset 2} around the Ca~II K line. The interstellar Ca~II K
  line is clearly visible in both spectra, shifted by about --30 \kms. Right panels show two
  spectra of {\sl Dataset 1} around the Mg~II line.}
\label{spec}
\end{figure}

For stars in {\sl Dataset 1} we measured \vrot\ from several different
spectral lines (typically 5-10, depending on the signal-to-noise
ratio), in order to have independent measures. Then the average value
is assumed as final measure.  For stars in {\sl Dataset 2}, RVs have
been measured from the Ca~II K line.  This feature is a robust
diagnostic for BSS rotational velocities, because it is strong enough
to be observable also in case of very high rotational velocity (while
weaker features, as those used for the spectra of {\sl Dataset 1}, can
be unobservable for \vrot\ $>$100 \kms). Moreover its strength
decreases when the surface temperature increases, making easier to
disentangle the contribution of the rotational velocity from the
intrinsic line profile.  However, among the coldest stars (where the
intrinsic width of the Ca~II K line can make the measurement of low
\vrot\ more uncertain) other transitions are detectable and have been
used to confirm the derived values of \vrot.  As discussed above, the
interstellar Ca~II K line has been masked in order to not affect the
fitting procedure of the photospheric line.  Fig.~\ref{spec} shows
examples of a few spectra in the Ca~II K line spectral region (left
panels) and the Mg~II lines (right panels), with low and high
rotational velocities. The interstellar Ca~II K line, shifted by about
$-30$ \kms, is clearly visible in the left panels.  

Spectra observed with gratings HR4, HR14A and HR15N have not been used
to determine the rotational velocities.  This is because the only
  available features in these gratings are the $H_{\alpha}$ and
  $H_{\gamma}$ Balmer lines, which can be poor diagnostics of \vrot,
  due to their intrinsic width, their high sensitivity to atmospheric
  parameters and their comparatively much lower sensitivity to
  rotation (see Sect. \ref{cfr}).

As quoted above, 17 stars have been found in common in the two
datasets. The comparison of the two independent \vrot\ measures
allows an important check of consistency. The average difference
between \vrot\ for stars in common between {\sl Dataset 1} and {\sl 2}
turns out to be quite small, $\Delta v_e= (v_e \sin i)_1 - (v_e \sin
i)_2 =-0.6 \pm 1.7$ \kms\ ($\sigma=6.7$ \kms).  Such a difference (consistent 
with zero \kms\ ) guarantees that no systematics are present in our 
analysis and its small rms uncertainty can be adopted as an estimate 
of the internal accuracy of our measurements. 
For stars for which two measurements of
\vrot\ are available, we assumed the mean rotational velocity. For two
stars in common between the two datasets for which only lower limits
are derived from the spectra of {\sl Dataset 1}, we assumed the value
obtained from {\sl Dataset 2}. Finally, 5 stars available only in {\sl
  Dataset 1} show featureless spectra for the HR5 grating, and hence
only lower limits can be derived by the comparison between observed
and synthetic spectra.  Finally, the external accuracy of our
  measures is demonstrated by the good agreement with the independent
  results of SP14 for 13 targets in common between the two datasets
  (see Sect. \ref{cfr}), and by the values we obtained for a sample
  of 24 sub-giant branch stars observed during the same runs discussed here (they
  are all consistent with zero \kms, in agreement with the typical
  rotational velocity of stars in this evolutionary phase).

%%%%% GRAVITY DARKENING
It is important to specify that the values of \vrot\  are derived
by assuming that the stellar atmospheric parameters do not vary with
the star latitude. This is not totally true for very rapidly rotating
stars, that are affected by the so-called gravity darkening
\citep{vonziepel}, usually expressed by a power law, $T_{\rm
  eff}\propto g_{\rm eff}^{\beta}$ (with $\beta\le 0.25$), implying
that the equatorial regions are cooler than the poles \citep[see
  also][]{espinosa}.  This effect should lead to an underestimate of
$\sim$10-20\% of \vrot\  for stars with relevant rotation ($>
200$ \kms), as O and B-type stars (see \citealp{town04} and
\citealp{ramir}). Our results do not include corrections for gravity
darkening; however, this effect should only marginally affect the
observed BSSs (since most of them rotate much more slowly than the
OB-type stars), thus not modifying our conclusions on the \vrot\ 
distribution.

Uncertainties in the fitting procedure of each individual line have
been estimated by resorting to Montecarlo simulations, creating for
some targets (representative in terms of atmospheric parameters and
spectral quality) a sample of 1000 synthetic spectra with added
Poissonian noise, and repeating the analysis. The
typical uncertainties for slow-rotating stars range from $\sim$3 \kms\  
up to $\sim$5 \kms, according to the signal-to-noise ratio.  For FR
stars the uncertainties can reach $\sim$20 \kms, in particular
for the BSSs with \vrot\  larger than $\sim$150 \kms.  For
the stars for which \vrot\  has been derived from different
lines, we assumed as internal uncertainty the dispersion of the mean
(weighted on the MonteCarlo uncertainties of each individual line)
normalized to the square root of the number of lines. For the stars
for which only one line has been used (as the majority of the targets
of {\sl Dataset 2}), the MonteCarlo uncertainty has been adopted.

Fig. \ref{distr} shows the distribution of \vrot\ for the entire BSS
sample, with the lower limits for the five BSSs also shown with
arrows.  The distribution turns out to be very broad, with a main peak
at low velocities (5-20 \kms) and a long tail toward high
\vrot\ values (reaching $\sim$200 \kms).  Note that such large
values are unusual and unexpected for {\it normal} GC stars, where the
highest rotational velocities reach only values of $\sim 40$-50
\kms\ and are observed among the Horizontal Branch stars hotter than
the instability strip and colder than the {\it Grundahl Jump}
\citep[see e.g.][]{behr00a,behr00b,lovisi13a}.  Thus, BSSs populating
the tail of the distribution certainly are the most FR stars observed
in GCs.  Instead, among the MS field stars the rotational velocity is
a function of the temperature \citep{cortes}, with \vrot\ $\sim 10$
\kms\ for stars colder than $\sim 9000$ K, while the
\vrot\ distribution covers a broad range, up to values larger than 200
\kms\ for hotter stars; note that the most rapidly rotating star
  discovered so far \citep[namely VFTS102 in 30 Doradus, ][]{dufton}
  is a O-type stars with a projected rotational velocity larger than
  500 \kms.  However, as shown in Fig. \ref{temp} such a behaviour is
not observed in our BSS sample.  In particular, all the BSSs with
\vrot\ larger than 100 \kms\ have $T_{\rm eff}$ lower than $\sim$8800
K, while for higher temperatures the BSSs show rotational velocities
lower than $\sim$90 \kms. This provides additional support to the fact
that BSSs do not behave as genuine MS stars.

\section{Comparison with SP14}
\label{cfr}

SP14 provide the rotational velocities of 47 BSSs in $\omega$ Centauri
measured by using IMACS@Magellan spectra (R$\sim$10000). In their
work, the derivation of \vrot\ is based on the penalized pixel
  fitting of Gauss-Hermite expansion of the line profile adopting as
templates 12 spectra of stars with spectral type similar to that
of the target BSSs, acquired with the ELODIE spectrograph at the
Observatoire de Haute-Provence. Fourteen of our targets are in common
with their sample.  Fig. \ref{compuz} shows the residuals between the
rotational velocities measured in the two works.  Apart from star
\#111709 (for which a large discrepancy of 140 \kms\ is found) and
star \#111909 (for which we provide only a lower limit), the average
difference is (\vrot)$-(v_{e}$sin i$)_{\rm SP14} =-6.3 \pm 4.3$ \kms,
with a quite large dispersion about the mean ($\sigma=14.8$ \kms).
The residuals show a quite well defined trend with large differences
(up to $\sim$35 \kms) for low rotational velocities (\vrot\ $<$20-30
\kms) and a substantial agreement at large velocities. This suggests
that the differences between the two works are essentially due to the
relative low resolution of IMACS (which is about half the FLAMES
one). As expected in fact at lower resolution is more difficult to
correctly measure small \vrot\, because the line profile is dominated
by the instrumental broadening.

  We investigated in detail the origin of the large difference in
  the rotational velocity of star \#111709 measured in the two
  studies: \vrot$=165$ \kms\ in SP14, \vrot$=29$ \kms\  in our work.  In
  Fig. \ref{mg} we show a portion of the spectrum of this star around
  the Mg~II line at $4481\mathring{A}$, with overimposed synthetic
  spectra calculated with the rotational velocity derived in our
  analysis (thin solid line) and in SP14 (thin dashed line). The
  comparison clearly demonstrates that the observed profile of the
  Mg~II line cannot be reproduced with the large value of
  \vrot\ quoted by SP14. We attribute such a discrepancy to the
  different lines used to measure the rotation velocity.

In fact, in the spectral ranges investigated by SP14 and at the
relatively low spectral resolution of IMACS, metallic transitions very
sensitive to rotation become weak or even disappear in hot stars like
BSS \#111709 ($T_{\rm eff}=10250$ K), and the only measurable lines
are those of the Balmer series.  In general, these lines are only
marginally sensitive to rotation and suffer from several effects not
easily modeled, nor well constrained.  In fact, the detailed
inspection of a grid of synthetic spectra computed with fixed
atmospheric parameters and different values of \vrot\ reveals that the
core of the line is the only region of the Balmer profile sensitive to
rotation.  As an example, in Fig. \ref{bal} we compare synthetic
spectra around the $H_{\gamma}$ and $H_{\beta}$ Balmer lines,
calculated by assuming the same atmospheric parameters
($T_{\rm eff}=10000$ K, logg$=4.5$) and by varying only the rotational
velocity.
As apparent, an increase of \vrot\ from 0 \kms\ up to 200 \kms,
produces no effects on the FWHM and the wings of the lines
\citep[which are instead very sensitive to $T_{\rm eff}$/log~g
  variations, as discussed by][]{fuhrmann}. The variations of the core
depth are relatively small with respect to the assumed changes of
\vrot, and only a few pixels of the entire line profile can therefore
be used. On the other hand, small variations of the atmospheric
parameters lead to large variations of the entire line profile.
Moreover, since the core of the Balmer lines forms in the most
external layers of the photosphere, it can suffer from several effects
which are not easy to take into account and that could vary from star
to star (e.g., departures from the Local Thermodynamical Equilibrium,
stellar winds, chromospheric activity, convective granulation).  All
these effects prevent the use of the Balmer lines as solid indicators
of rotational velocity. In general, they only allow to roughly
discriminate between slow and very fast rotating stars, with large
uncertainties in the derived values of \vrot\ (as clearly visible in
Fig. \ref{bal}).

At the temperature and gravity of BSS \#111709, the Balmer lines show
very large wings, and metallic feature sensitive to
rotations could be hardly measurable, especially at moderate spectral
resolution. This is the most likely explanation for the
large discrepancy found between the value of \vrot\ estimated by SP14
(from the Balmer lines), and the one we obtained from some metallic lines. 
Because of their lower effective
temperatures, such a problem is significantly mitigated for the
remaining stars in common between the two studies (see the inset in
Fig. \ref{compuz}).  We note that for the vast majority of the targets
in $\omega$ Centauri, SP14 measured the rotational velocity from the
spectral region around the $H_{\gamma}$ and $H_{\beta}$ Balmer lines,
and only for a small sub-set they also used the spectral regions
around the Mg~II line at $4481\mathring{A}$ and the Mg {\it b} triplet
at $\sim 5180 \mathring{A}$.  
Thus, the rotational velocity of the hottest stars could be biased, if 
only Balmer lines were used in the analysis. However, given 
the general agreement between the
two samples (Fig. \ref{compuz}), in the next section we also
discuss the two datasets together (for the stars in common we adopted
our FLAMES values because of the higher spectral resolution).

\section{Discussion and Conclusions}
 
In this work we presented the rotational velocities of 109 BSSs in
$\omega$ Centauri.  This is the largest homogeneous characterization
of the rotational properties of the BSS population ever performed in a
single stellar system. 

\begin{figure}
\plotone{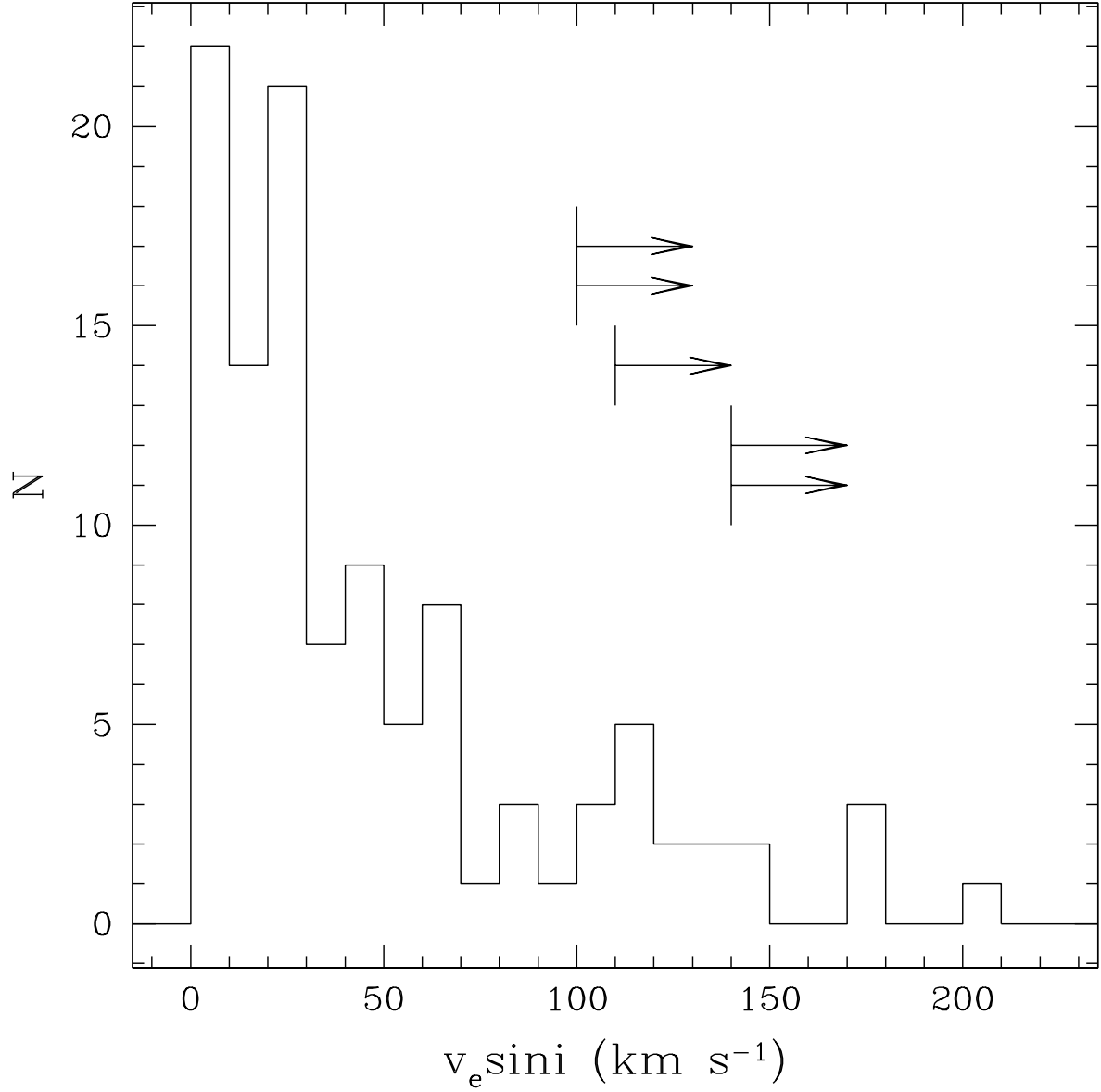}
\caption{Rotational velocity distribution for the 109 BSSs studied in
  $\omega$ Centauri. In five cases only lower limits to
  \vrot\ (arrows) could be derived.}
\label{distr}
\end{figure}

\begin{figure}
\plotone{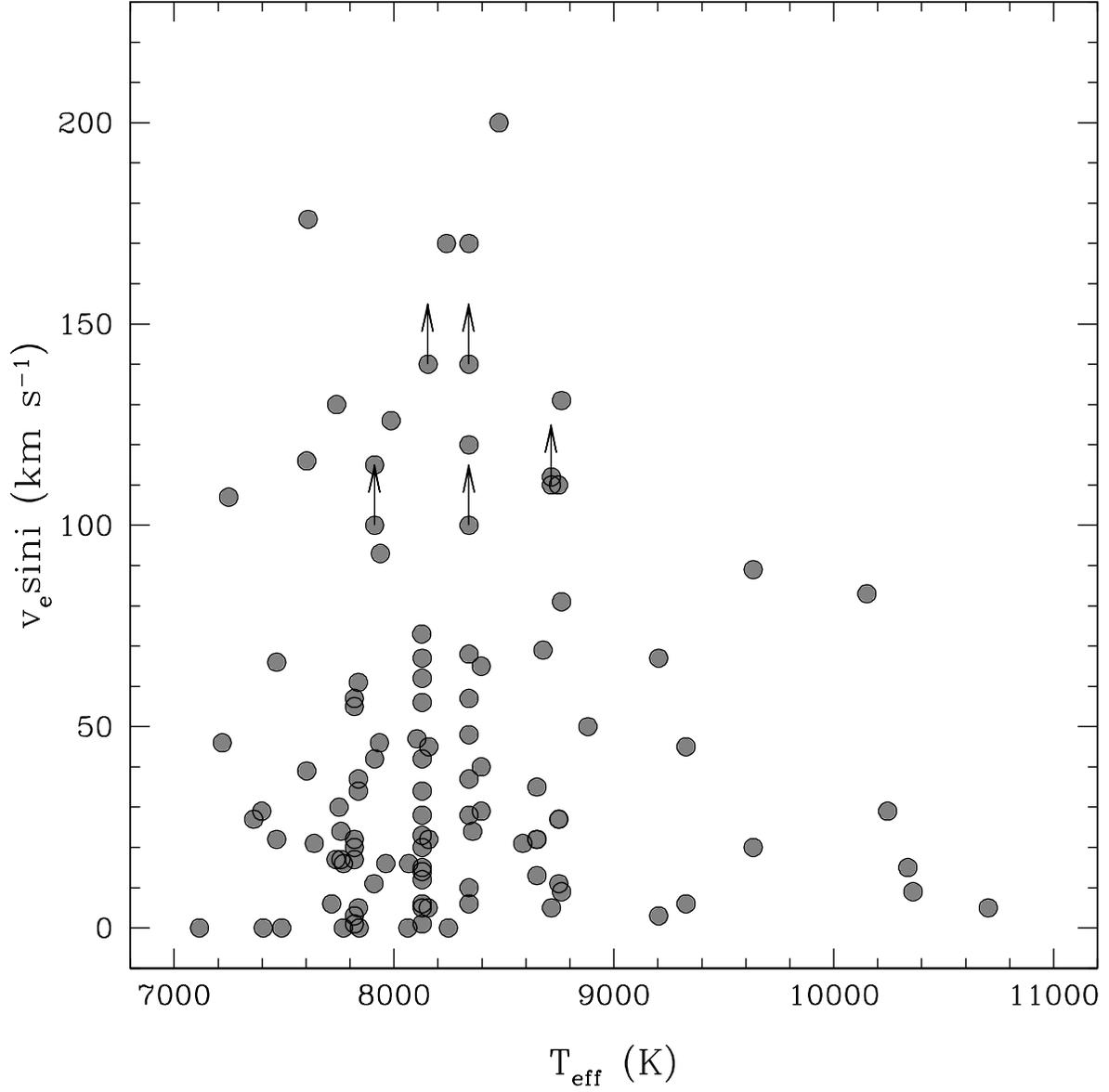}
\caption{Behaviour of the projected rotational velocity as a function
  of the effective temperature.  Arrows indicate \vrot\  lower
  limits.}
\label{temp}
\end{figure}

\begin{figure}
\plotone{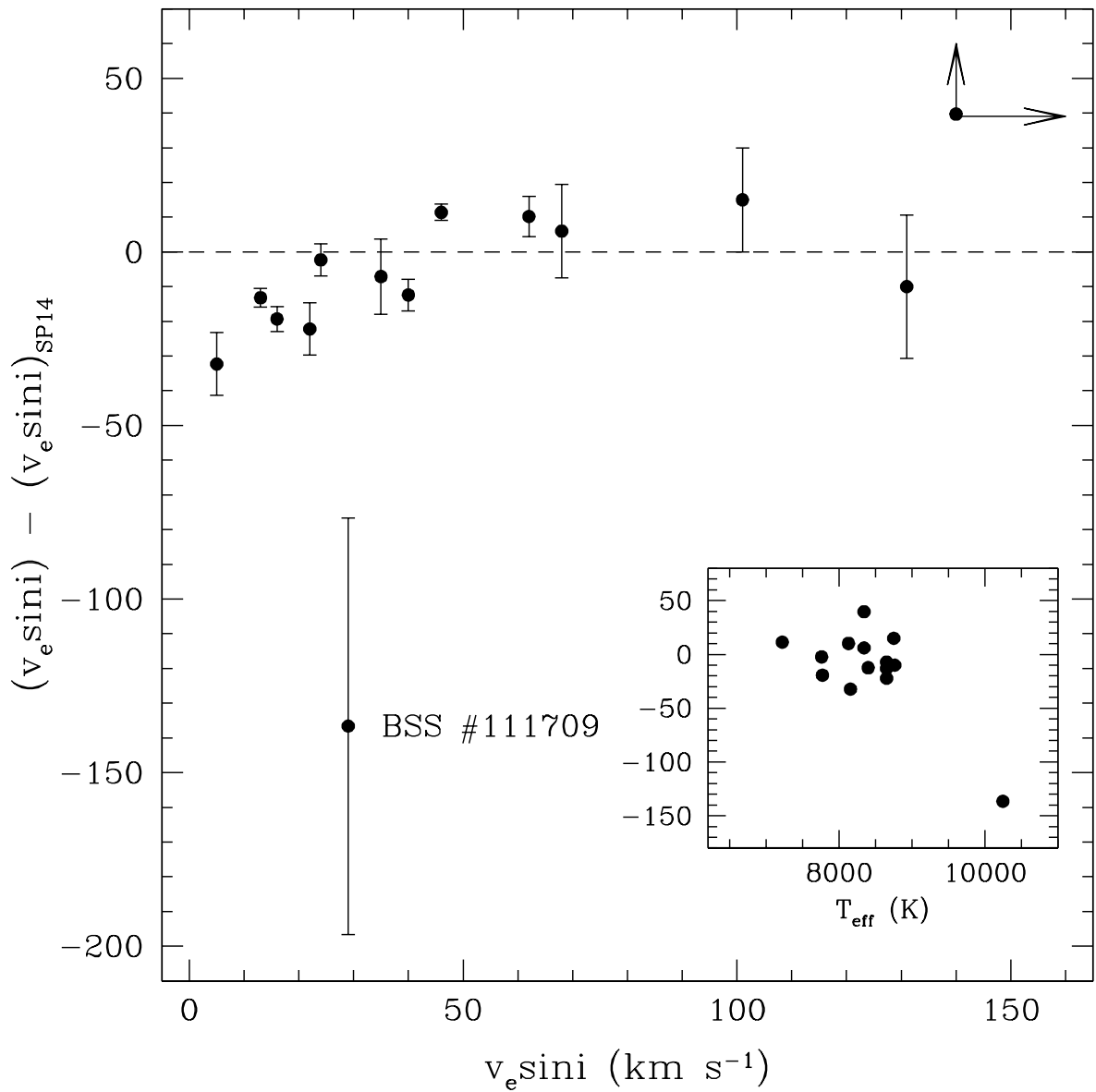}
\caption{Difference between the rotational velocities measured in this
  work and those of SP14, as a function of our \vrot\ for the 14 BSSs
  in common between the two samples. Errorbars are calculated as the
  sum in quadrature of the individual uncertainties. Note that the
  large errorbar of the BSS \#111709 is due to the large uncertainty
  quoted by SP14 ($\sim 60$ \kms).  The inset shows the
    rotational velocity differences as a function of the stellar
    effective temperature.}
\label{compuz}
\end{figure}

\begin{figure}
\plottwo{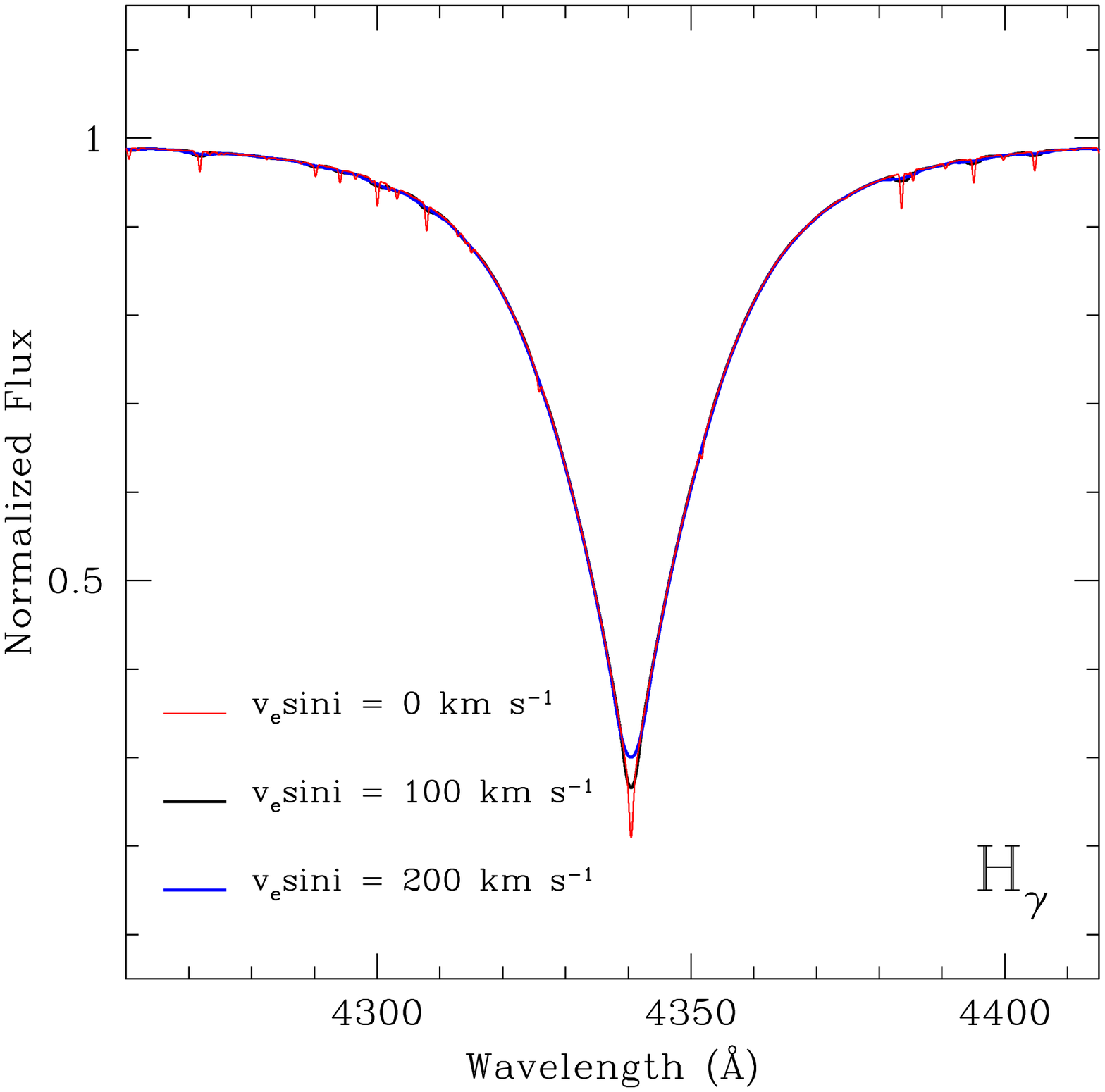}{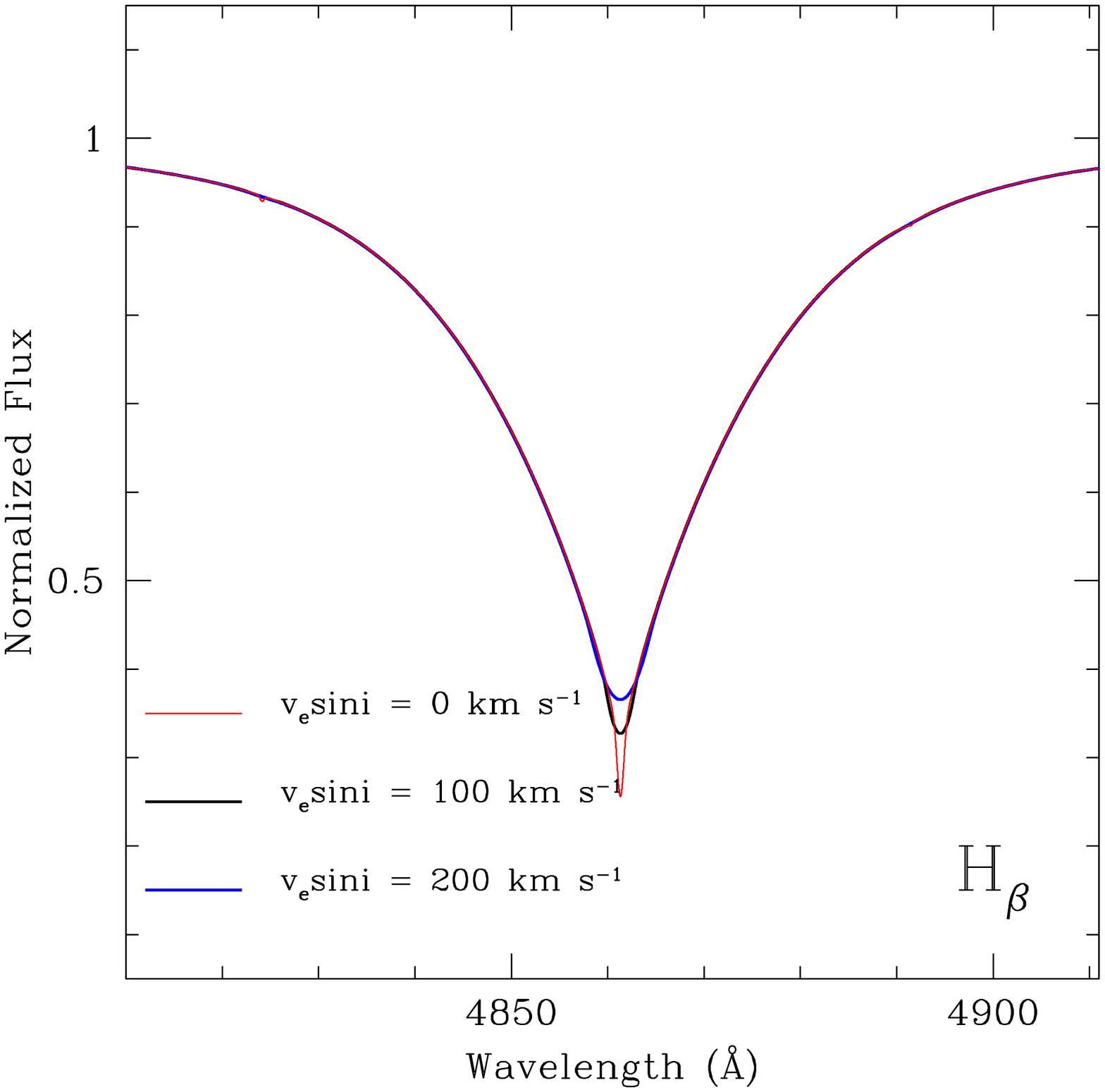}
\caption{Comparison between synthetic spectra calculated by
    assuming $T_{\rm eff}=10000$ K, logg$= 4.5$ and [M/H]$=-1.5$ dex,
    and different values of the rotational velocity (\vrot $=0, 100,
    200$ \kms) for the spectral regions around the $H_{\gamma}$ and
    $H_{\beta}$ (left and right panel, respectively).}
\label{bal} 
\end{figure}

\begin{figure}
\plotone{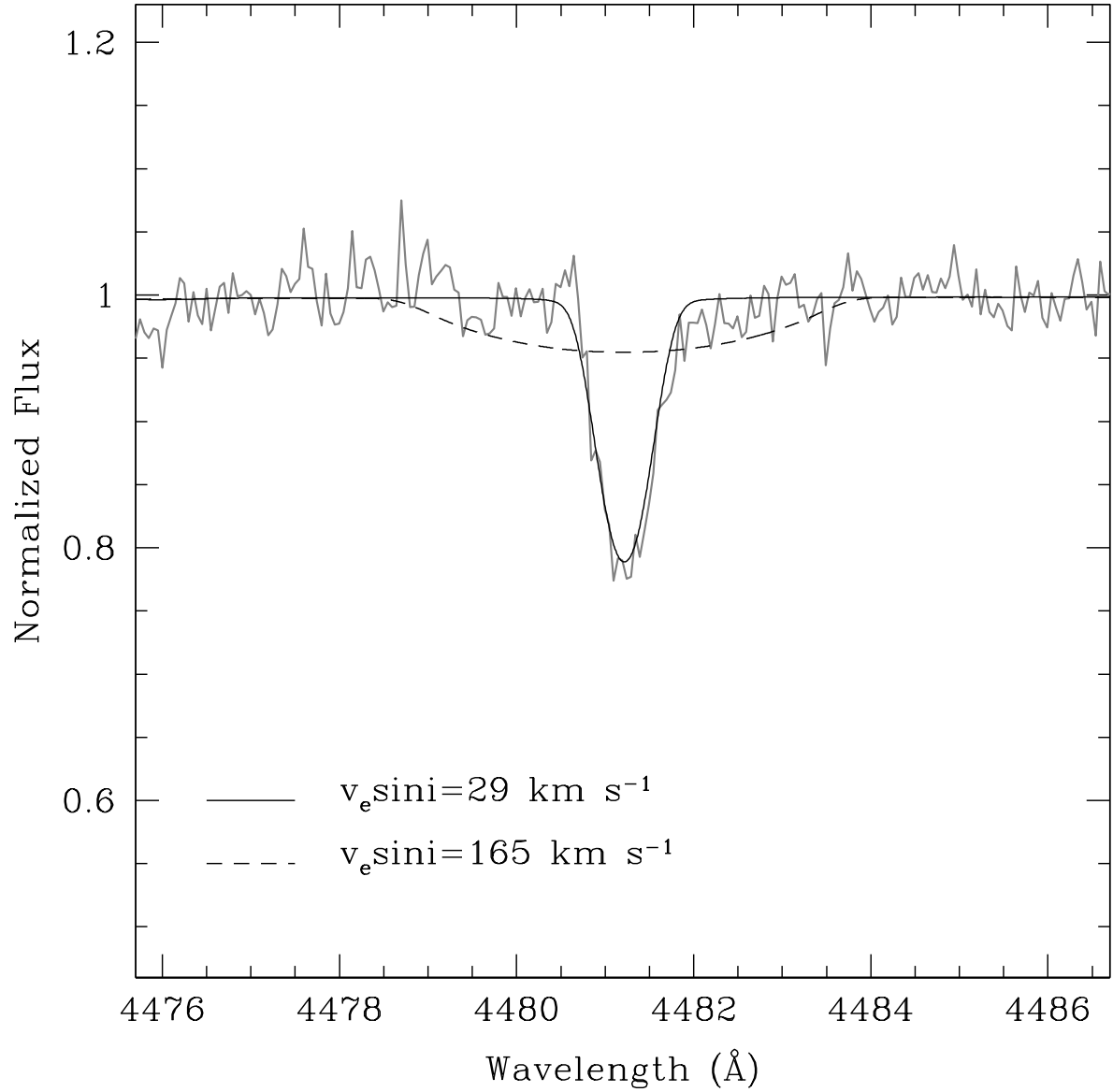}
\caption{Portion of the spectrum around the Mg~II line for BSS
  \#111709, one of the targets in common with the dataset of
  SP14. Overimposed are two synthetic spectra calculated with the
  rotational velocity derived in this work (solid black line) and in
  SP14 (dashed black line). }
\label{mg}
\end{figure}

\begin{figure}
\plotone{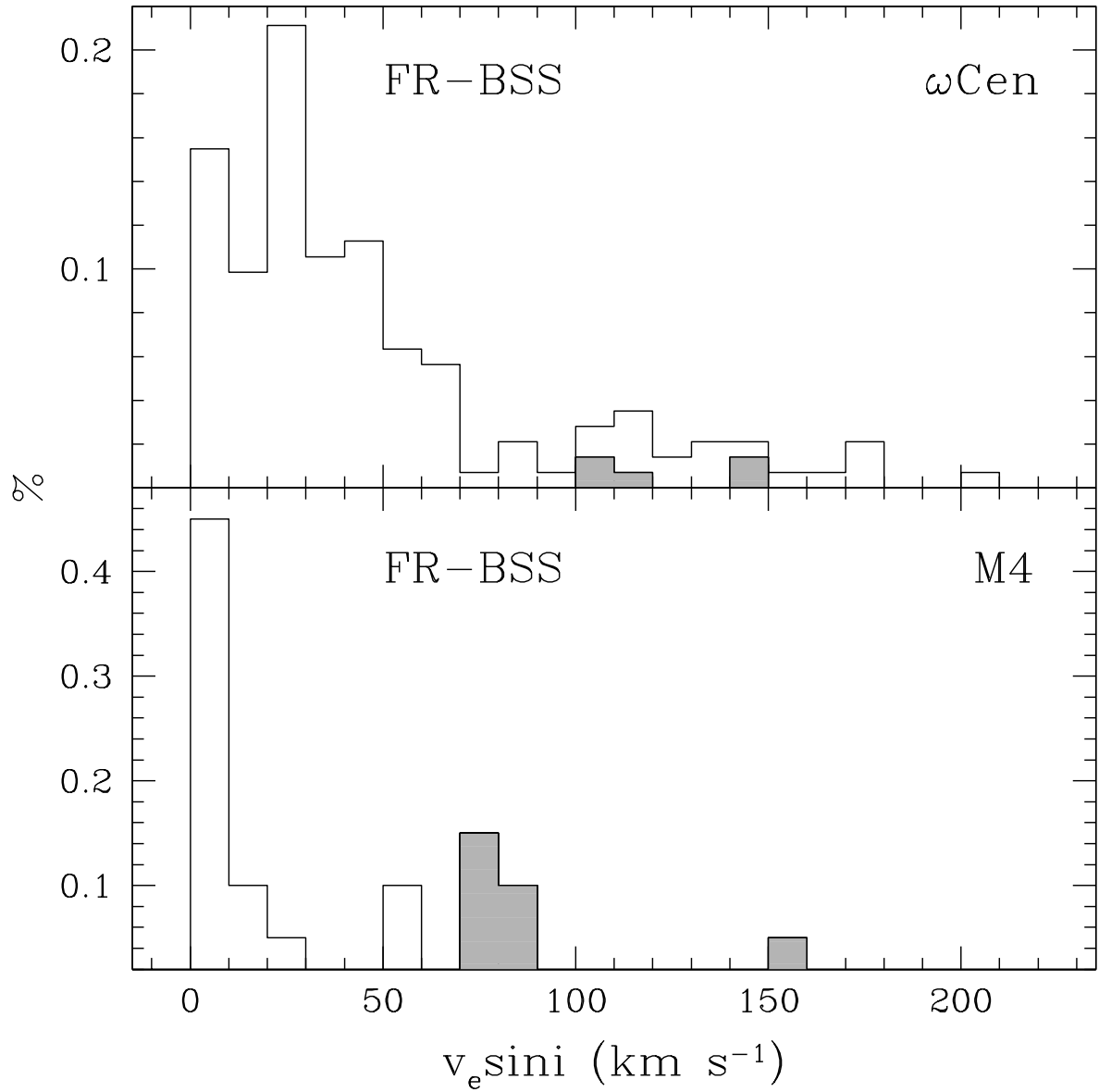}
\caption{\emph{Upper panel}: rotational velocity distribution for the
  combined sample of BSSs in $\omega$ Centauri (109 stars studied
    in this work with FLAMES spectroscopy and 33 targets from
    SP14). \emph{Lower panel:} rotational velocity distribution for
    the 20 BSSs studied by \citet{lovisi10} in M4. Grey histograms
    refer to upper limits.}
\label{omega_m4}
\end{figure}

As shown in Fig. \ref{distr}, the distribution of rotational
velocities is peaked at low values (10-30 \kms), with a long tail
toward larger values. Indeed the fraction of BSSs populating the tail
is considerable. By assuming \vrot$ = 40$ \kms\ as a reasonable
threshold between slow and fast rotators (see also Lovisi et
al. 2010), it turns out that 44 BSSs (out of the 109 observed) are
FRs. This corresponds to $\sim 40$\% of the studied sample. 
  Adding the BSSs analysed in SP14 and not in common with the FLAMES
  dataset, we obtain a total of 142 BSSs.  The rotational velocity
  distribution of the combined sample is shown in the upper panel of
  Fig. \ref{omega_m4}. The overall shape well corresponds to that in
  Fig. \ref{distr}, and also the fraction of BSSs spinning faster
  than 40 \kms\ remains unchanged (59 BSSs are FRs, corresponding to
  $\sim$40\% of the entire sample).  However, the distribution of
  \vrot\ shown in Fig. \ref{omega_m4} suggests that a higher velocity
  boundary for separating FRs may be more appropriate.  Consistently
  with SP14, we thus assumed 70 \kms\ as a threshold value, finding
  that 28 out 142 BSS (20\%) can be labelled as fast spinning
  BSSs. This is in agreement with the fraction obtained by considering
  our FLAMES sample alone, and with that quoted by SP14 for their
  dataset. Interestingly enough, the fraction of FR BSSs in M4
  (\citealp{lovisi10}; see lower panel of Fig. \ref{omega_m4} for a
  direct comparison) is very similar to that of $\omega$ Centauri,
  independently of the adopted boundary ($\sim$40\% with \vrot=40
  \kms, and $\sim$30\% with 70 \kms), even though the total size of
  the sample is considerably smaller (including only 20 stars).
Totally different results are obtained for the other GCs investigated
so far, namely 47 Tuc, NGC 6397, M30 and NGC 6752
\citep[][respectively]{f06_tuc, lovisi12, lovisi13a, lovisi13b}, where
the bulk of BSSs has \vrot$ < 30-40$ \kms\ and only one star per
cluster (if any) has high rotational velocity (larger than
$\sim$80-100 \kms).  Thus, besides M4, $\omega$ Centauri is the second
cluster with a significant fraction of FR BSSs.

\begin{figure}
\plotone{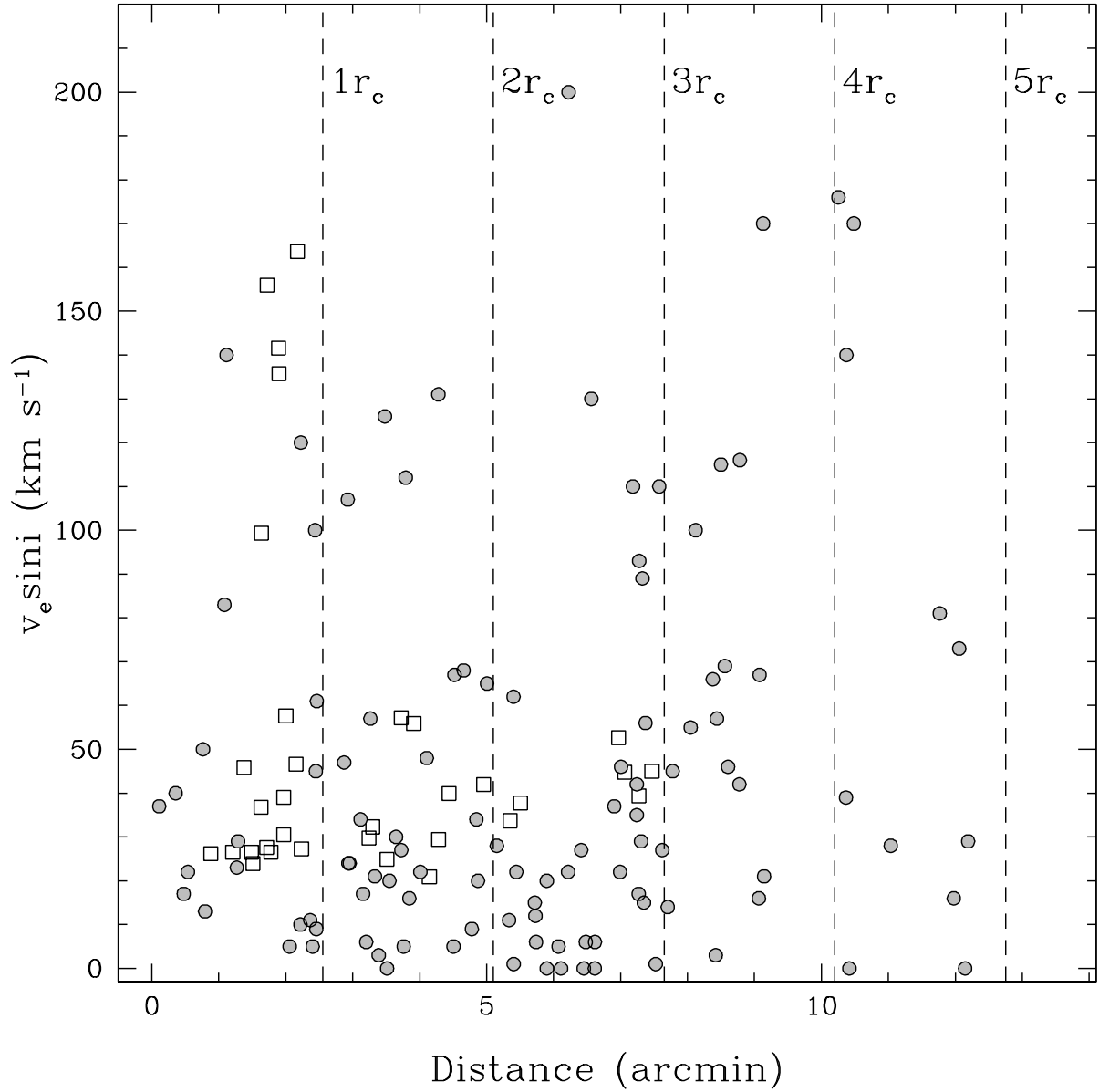}
\caption{Rotational velocities for the combined BSS sample in
    $\omega$ Centauri, as a function of the distance from the cluster
    center: FLAMES targets (grey circles) and BSSs from SP14 (empty
    squares). The vertical dashed lines mark the distance from the
    cluster center in units of the core radius.}
\label{dist_fr}
\end{figure}

\begin{figure}
\plotone{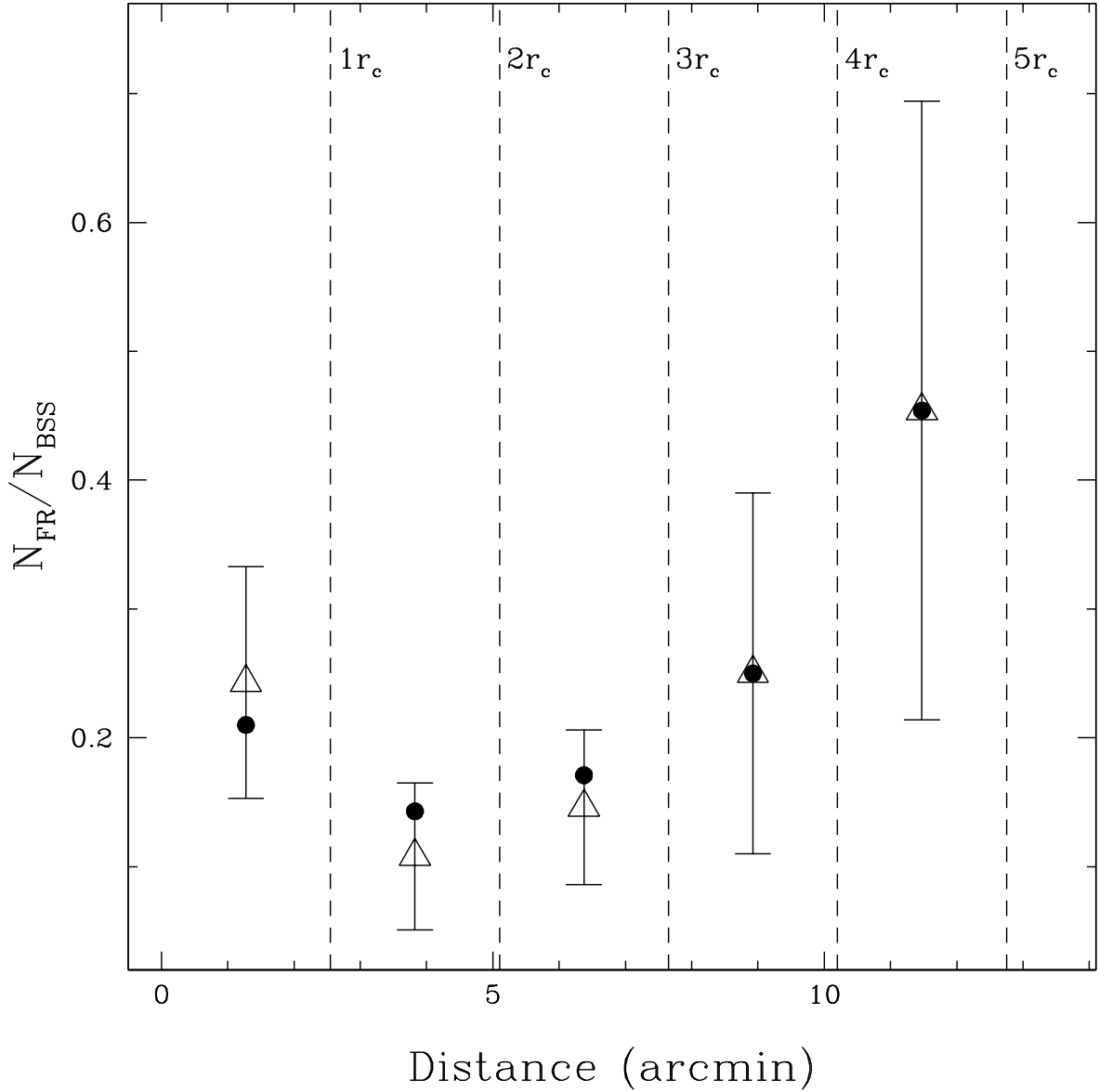}
\caption{ Number ratio between fast rotators and all BSSs, as a
    function of the distance from the cluster center. Black circles
    correspond to the FLAMES dataset alone, empty triangles to the
    combined BSS sample. Errorbars are referred to the combined BSS sample.}
\label{dist_ratio}
\end{figure}

Interesting insights can be drawn also from the BSS spatial
distribution.  SP14 found that the FR-BSSs (with \vrot$>70$ \kms) in
$\omega$ Centauri are more segregated toward the center, with respect
to the slowly rotating objects.\footnote{ We note that the
    distances quoted in SP14 have been calculated by neglecting the
    term cos(Dec). By taking this term into account and computing the
    correct distances, we find that all their targets are located
    within 3$r_c$ (instead of $\sim4.6r_c$) from the center, and all
    the FR-BSSs but two (i.e., 7 out of 9) are positioned within $1
    r_c$ (instead of $2 r_c$).}  While the targets in their
  sample cover only the inner $\sim 8\arcmin$, the combined dataset
  extends out to $\sim 13\arcmin$, corresponding to $\sim 4.8 r_c$
  (the adopted cluster center and core radius are from
  \citealp{f06_ome}: $\alpha_{J2000} = 13^h$ $26^m$ $46.5^s$,
  $\delta_{J2000}=-47^\circ$ $28\arcmin$ $41.1\arcsec$, $r_c=2.55
  \arcmin$).  The distribution of \vrot\ as a function of the distance
  from the cluster center is shown in Fig. \ref{dist_fr} for our
  sample (grey circles) and for the additional targets from SP14 (empty
  squares). The ratio between the number of FRs and the total number of
  BSSs ($N_{\rm FR}$/$N_{\rm BSS}$), measured in concentric annuli
  with radii equal to 1, 2, 3, 4 and 5 $r_c$, is plotted in
  Fig. \ref{dist_ratio}. Our measurements alone (solid circles), as
  well as the combined sample (empty triangles), show a trend
  with the distance from center.  In agreement with the findings of
  SP14, the fraction of FRs is larger in the central regions than at $\sim 3 r_c$ 
  ($N_{\rm FR}/N_{\rm BSS} = 0.24$ and 0.11, respectively for the combined sample). 
 A larger value is found in the
    most external regions, beyond $4 r_c$, where the fraction of fast
    spinning BSSs reaches the 45\% of the total.  
  Because of the small number statistics of the sample (in particular 
  for the outermost regions), the uncertainty in each single bin is large.
  In order to assess the statistical significance of this possible behaviour, 
  we performed a Kolmogorov-Smirnov test between the radial distributions 
  of the populations  of fast- and slow-rotating BSSs, yielding a 
  $\sim$20\% probability that the two samples are extracted from the same 
  parent distribution.
  Even if this test provides a statistical significance smaller than 2$\sigma$, 
  a different radial distribution for FR and slow-rotating BSSs cannot be 
  totally ruled out.
  Also, we checked this result against changing the coordinates of the
    cluster center according to the values quoted in the literature by
    \citet{ander}, \citet{vanlew} and \citet{noyola}, finding no
    variations in the observed behavior.  On the other hand, no
    similar trend is observed for M4, the other GC with a large
    population of FR BSSs. However, the size of the sample is very
    limited in that case and we cannot therefore exclude that also in
    M4 the fraction of fast rotating BSSs has a bimodal radial trend,
    with a significant rise outward.

Based on the flat BSS radial distribution observed in $\omega$
Centauri, \citet{f06_ome} suggested that the vast majority of these
stars (if not all) has a non-collisional origin and probably formed
through MT in binary systems.  On the other hand, theoretical models
predict high rotational velocities for MT-BSSs, at least at the moment
of their formation \citep{sarna,lombardi}. The occurrence of
subsequent processes, like magnetic braking or disk locking, might
then slow down the stars.  Thus it is reasonable to conclude that the
high rotational velocity observed in the FR BSSs of $\omega$ Centauri
is the observational confirmation that significant transfer of angular
momentum occurs during the MT-BSS formation process.  An additional
support comes from the fact that all the 6 W Uma stars in our sample
are FR BSSs (with \vrot\ $> 90$ \kms).  These variables are compact
binary systems experiencing an active phase of MT, and their evolution
is thought to lead to the total coalescence of the two components,
eventually forming a BSS \citep{vilhu}.  Note that W UMa stars with
high rotational velocities have been found also in 47 Tuc
\citep{f06_tuc}, M4 \citep{lovisi10} and M30 \citep{lovisi13b}. 
{Only two W UMa stars with slow rotational velocities have been observed
  (in 47 Tuc; see \citealp{f06_tuc}), and both show evidence of
  CO-depletion on their surfaces, which is interpreted as the chemical
  signature of the MT process.  This might suggest that these stars
  have been caught when they are still accelerating toward very high
  rotational velocities: enough CO-depleted material was already
  transferred onto the secondary to be detectable in its atmosphere
  and the star rotational velocity is not yet too high to prevent the
  spectroscopic measurement of chemical abundances.}  Under the
assumption that most of the BSSs in $\omega$ Centauri are generated by
MT, the discovered FRs could either be recently formed BSSs (for which
the braking mechanisms have not had enough time yet to be effective),
or, for some unknown reason, they could have been able to preserve
longer their initial high rotation.  Indeed, no significant
distinction between fast and slowly rotating BSSs pointing toward a
different formation epoch has been observed in the CMD. On the other
hand, the detailed efficiencies and time-scales of the braking
mechanisms are still not known and therefore firm conclusions are hard
to be drawn here.  However, the observed radial distribution of
  fast and slow rotating BSSs, with a possible central peak and a
  significant outward rise of the fraction of fast spinning BSSs, seems
  to suggest that environment can play a role in this game.  In
  particular, it could indicate either that recently formed MT-BSSs
  prevail in the outskirts of $\omega$ Centauri, or that braking
  processes are least efficient in the lowest density regions of the
  system. The slight increase observed within $1 r_c$ may suggest that
  stellar interactions can contribute to transfer angular momentum and
  accelerate BSSs (see also \citealp{lovisi10} and SP14).  So far,
such a distribution has been detected only in $\omega$ Centauri and
further determinations of rotational velocities for large samples of
BSSs in other clusters are needed to provide a solid answer about the
ubiquity of this phenomenon.

\acknowledgements 
We thanks the anonymous referee for his/her comments and suggestions.
This research is part of the project COSMIC-LAB
(http://www.cosmic-lab.eu) funded by the European
Research Council (under contract ERC-2010-AdG-267675).  L. Lovisi
wishes to thank ESO for the hospitality at ESO-Santiago provided
during the preparation of the paper.

{}

%%%%%%%%%%%%%%%%%%%%%%% %%%%%%%%%%%% FIGURES %%%%%%%%%%%%%%%%%%%%%%%%%%%%%%%%%%%%%%%%%%%%%%%%%%%%

\begin{landscape} 
\begin{deluxetable}{ccccccccc}
\tablecolumns{9} 
\tablewidth{0pc}  
\tablecaption{Parameters and velocities of the observed BSS sample}
\tablehead{
\colhead{ID} &      Ra  &   Dec     & \colhead{$T_{\rm eff}$} & \colhead{$\log g$} & RV  & \vrot\   & Dataset & Variable\\
             &  (J2000) & (J2000)   &        (K)	    &		      &  (\kms)  &  (\kms) &   & }
\startdata 
\hline
  
  6365	&  201.6079938	 & -47.4235140   &   8762   &   4.33  &   221.8   &	9  &   1       &       ---     \\
  6452  &  201.6147225	 & -47.4540869   &   7770   &   4.29  &   220.7   &	0  &   2       &       ---     \\
 15741	&  201.6117617	 & -47.4889112   &   9204   &   4.15  &   214.6   &	3  &   1       &       ---     \\
 15822	&  201.5894328	 & -47.4673384   &   8762   &   4.33  &   226.4   &   131  &   1       &       ---     \\
 15913	&  201.5705488	 & -47.4571373   &   8128   &   4.29  &   226.7   &    28  &   1       &     V217 (SX)      \\
 25168  &  201.6458578	 & -47.5029603   &   8158   &   4.06  &   238.9   &    45  &   1       &       ---     \\
 25256	&  201.6394925	 & -47.4735261   &   8341   &   4.16  &   241.3   &    10  &   1       &       ---     \\
 25497	&  201.6330705	 & -47.4759309   &   7838   &   4.15  &   225.6   &    61  &   1       &       ---     \\
 45771	&  201.6301436	 & -47.5559363   &   7909   &   3.89  &   229.6   &    11  &   1--2    &       ---     \\
 45842	&  201.6315878	 & -47.5197093   &   9634   &   4.46  &   231.9   &    20  &   1       &       ---     \\
 45881	&  201.6323122	 & -47.5325120   &   8341   &   4.16  &   229.0   &    48  &   1--2    &    NV310  (SX)     \\
 45952	&  201.6438653	 & -47.5175900   &   8128   &   4.29  &   244.6   &    34  &   1       &     V198  (SX)    \\
 45972  &  201.6188194	 & -47.5173629   &   7770   &   4.29  &   216.1   &    16  &   2       &    NV319  (SX)    \\
\hline		 
\enddata 	 
\tablecomments{Identification number, coordinates, temperature,
  gravity, radial and rotational velocities, name of the spectroscopic
  Dataset and variable type for the observed BSSs with the corresponding name in the catalogs 
  by \citet{kal04} and \citet{kal05}. A complete version of
  the table is available in electronic form.}  

\end{deluxetable}
\end{landscape}

\end{document}